\documentstyle[12pt,axodraw,graphicx]{article}
\begin{document}

\begin{center}

{\bf {PLANCKIAN
SCATTERING BEYOND THE EIKONAL APPROXIMATION IN THE QUASI-POTENTIAL  APPROACH}}\\

{\large {\bf \small {\sc{Nguyen Suan Han}}} and {\bf \small
{\sc{Nguyen Nhu Xuan }}}$\footnote { e-mail: Lienbat76@gmail.com
\bf{ Submitted to European Physical Journal C.}} $ } \vspace*{0.3cm}

\end{center}

\centerline{Department of Theoretical Physics,Vietnam National
University, } \centerline{P.O.Box 600, BoHo, HaNoi  10000, Vietnam }
\centerline{20th April, 2008}


\begin{abstract}
\small { Asymptotic behavior of the scattering amplitude for  two
scalar particles by scalar, vector and tensor exchanges  at high
energy and fixed momentum transfers is reconsidered in quantum field
theory. In the framework of the quasi-potential approach and the
modified perturbation theory a systematic scheme of finding the
leading eikonal scattering amplitudes and its corrections are
developed and constructed.The connection between the solutions
obtained by quasi-potential and functional approaches is also
discussed.The first correction to leading eikonal amplitude is found
.

}
\end{abstract}


\section{Introduction} .

Eikonal scattering amplitude  for the high energy of the two
particles in the limit of high energies and fixed momentum transfers
is found by many authors in quantum field theory $[1-9]$, including
the quantum gravity in recent years $[9-20]$. Comparison of the
results obtained by means of the different approaches for this
problem has shown that they all coincide in the leading order
approximation, while the corrections (non-leading terms) provided by
them are rather different $ [15,17,20,21,22,23]$. Determination of
these corrections to gravitational scattering is now open problem
$[10-14]$. These corrections play a crucial role in  such problems
as strong gravitational forces near black hole, string modification
of theory of gravity and some other effects of quantum gravity $[9-20]$.\\

The purpose of the present paper is to develop a systematic scheme
based on modified perturbation theory to find the correction terms
to the leading eikonal amplitude for high-energy scattering by means
of solving the Logunov-Tavkhelidze quasi-potential equation
$[24-27]$. In spite of the lack of a clear relativistic covariance,
the quasi-potential method keeps all information about properties of
scattering amplitude which could be received starting from general
principle of quantum field theory $[25]$. Therefore, at high
energies one can investigate analytical properties of the
scattering, its asymptotic behavior and some regularities of a
potential scattering etc. Exactly, as it has been done in the usual
S-matrix theory $[24]$. The choice of this approach is dictated also
by the following reasons: 1. in the framework of the quasi-potential
approach the eikonal amplitude has a rigorous justification in
quantum field theory [4]; 2. in the case of smooth potentials, it
was shown that a relativistic quasi-potential and the Schrodinger
equations lead to
qualitatively identical results $[28,29]$.\\

The outline of the paper is as follows. In the second section the
Logunov-Tavkhelidze quasi-potential equation is written in the
operator form. In the third section  the solution of this equation
is presented in the exponent form which is favorable to modify the
perturbation theory. The asymptotic behavior scattering amplitude at
high energies and fixed momentum transfers is considered in the
fourth section. The lowest-order approximation of the modified
theory is the leading eikonal scattering amplitude. Corrections to
leading eikonal amplitude are also calculated. In the fifth section
the solution of quasi-potential equation is presented in the form of
a functional path integral. The connection between the solutions
obtained by quasi-potential and functional is also discussed. It is
shown that the approximations used are similar and the expressions
for correction to the leading eikonal amplitude are found identical.
Finally, we draw our conclusions.

\section{Two particle quasi-potential equation }

For simplicity, we shall first consider the elastic scattering of
two scalar nucleons interacting with  a scalar meson fields the
model is described by the interaction Lagrangian
$L_{int}=g\varphi^{2}(x)\phi(x)$. The results will be generalized to
the case of scalar nucleons interacting with a neutral vector and
graviton fields later. Following Ref.$[23]$ for two scalar particle
amplitude the quasi-potential equation with local quasi-potential
has the form:

$$T({\bf{p}},{\bf{p}'};s)=gV({\bf{p}}-{\bf{p}'};s)
+g\int  d{\bf{q}} V({\bf{p}}-{\bf{q}};s) K({\bf{q}}^{2},s)
T({\bf{q}},{\bf{p'}};s), \eqno(2.1)$$ \\
where
$K({\bf{q}}^{2},s)=\frac{1}{\sqrt{q^2+m^2}}\frac{1}{q^2+m^2-\frac{s}{4}-i\varepsilon}
$, $s=4({\bf{p}}^{2}+m^{2})=4({\bf{p'}}+m^{2})$ is the energy and
${\bf{p}}, {\bf{p'}}$ and are the relative momenta of particles in
the center of mass system in the initial and final states
respectively. Equation $(2.1)$ is one of the possible
generalizations  of the Lippman-Schwinger equation for the case of
relativistic quantum field theory. The quasi-potential $ V $ is a
complex function of the energy and the relative momenta. The
quasi-potential  equation simplifies considerably if $ V $ is a
function of only the difference of the relative momenta and the
total energy, i.e. if the quasi-potential is local \footnote{Since
the total energy $ s $ appears as an external parameter of the
equation, the term "local" here has direct meaning and it can appear
in a three-dimensional $\delta $-function in the quasi-potential in
the coordinate representation }. The existence of a local
quasi-potential has been proved rigorously in the weak coupling case
$[27]$ and a method has been specified for constructing it. The
local potential constructed in this manner gives a solution of Eq.
$(2.1)$, being equal to the physical amplitude on the mass shell $ [24-26]$ .\\

Making the following Fourier transformations

$$
V({\bf{p}}-{\bf{p'}};s)=\frac{1}{(2\pi)^{3}}\int d{\bf{r}}
e^{i({\bf{p}}-{\bf{p}'}){\bf{r}}} V({\bf{r}};s),\eqno (2.2)$$
$$
T({\bf{p}},{\bf{p'}};s)=\int d{\bf{r}} d {\bf{r'}}
e^{i({\bf{p}}{\bf{r}}-{\bf{p'}}{\bf{r'}})} T({\bf{r}},{\bf{r'}};s).
\eqno(2.3)$$

Substituting $(2.2)$ and $(2.3)$ into $(2.1)$ , we obtain

$$
T(\bf r, \bf r' ; s)=\frac{g}{(2\pi)^{3}}V(\bf r; s)\delta^{(3)}(\bf
r- \bf r')+ $$
$$\frac{g}{(2\pi)^{3}}\int\int d\bf q K( \bf q^{2}; s)
V(\bf r; s) e^{-\bf q \bf r} \int d\bf r'' e^{i\bf q \bf r''} T( \bf
r'', \bf r' ; s) \eqno (2.4)
$$
and introducing the representation

$$
T({\bf{r}},{\bf{r}'};s)= \frac{g}{(2\pi)^{3}}
V({\bf{r}};s)F({\bf{r}},{\bf{r'}};s), \eqno(2.5)$$
we obtain

$$
F({\bf{r}},{\bf{r'}};s)=\delta^{(3)}
({\bf{r}}-{\bf{r'}})+\frac{g}{(2\pi)^{3}}\int d{\bf{q}}
K({\bf{q}}^{2};s) e^{-i{\bf{q}}{\bf{r}}}\times
$$

$$ \times \int
d {\bf{r''}}e^{i{\bf{q}}{\bf{r''}}}
V({\bf{r''}};s)F({\bf{r''}},{\bf{r'}};s). \eqno(2.6)
$$

Defining the pseudo-differential  operator

$$
\widehat{L_{r}}=K( - {\bf{{\nabla_{r}}}}^{2};s), \eqno(2.7)
$$

then

$$
K({\bf{r}};s)=\int d{\bf{q}}
K({\bf{q}}^{2};s)e^{-i{\bf{q}}{\bf{r}}}=K( -\nabla_{r}; s) \int
d{\bf{q}} e^{-i{\bf{q}}{\bf{r}}} = \widehat{L_{r}}
(2\pi)^{3}\delta^{(3)}({\bf{r}}). \eqno(2.8)
$$

With allowance for $(2.7)$ and $(2.8)$, Eq. $(2.6)$ can be rewritten
in the symbolic form:

$$
F({\bf{r}},{\bf{r'}};s)=\delta^{(3)}({\bf{r}}-{\bf{r'}})
+g\widehat{L_{r}}\bigl[V({\bf{r}},s)F({\bf{r}},{\bf{r'}},s)\bigl].
\eqno(2.9)
$$

Eq. $(2.8)$ is the Logunov-Tavkhelize quasi-potential equation in
the operator form.

\section{ Modified perturbation theory}

In the framework of the quasi-potential equation the potential is
defined as an infinite power series in the coupling constant which
corresponds to the perturbation expansion of the scattering
amplitude on the mass shell. The approximate equation has been
obtained only in the lowest order of the quasi-potential. Using this
approximation the relativistic eikonal expression of elastic
scattering amplitude was first found in quantum field theory for
large energies and fixed  momentum transfers $[22]$. In this paper
we follow a somewhat different approach based  on the idea of the
modified perturbation theory proposed by Fradkin $[30]$
\footnote{The interpretation of the perturbation theory from the
view-point of the diagrammatic technique is as follows. The typical
Feynman denominator of the standard perturbation theory is of the
form $(A)$: $(p+\sum q_i )^{2}+m^{2}-i\varepsilon=
p^{2}+m^{2}+2p\sum q_i +(\sum q_i)^{2}$, where $ p $ is the external
momentum of the scalar (spinor) particle, and the $ q_i$ are virtual
momenta of radiation quanta. The lowest order approximation $ (A) $
of modified theory is equivalent to summing all Feynman diagrams
with the replacement: $ (\sum q_i)^{2}=\sum (q_i)^{2} $ in each
denominator $(A)$. The modified perturbation theory thus corresponds
to a small correlation of the radiation quanta: $
\bf{q_{i}}\bf{q_{j}}=0$ and is often called the
$\bf{q_{i}}\bf{q_{j}}$-approximation. In the framework of functional
integration this approximation is called the straight-line path
approximation i.e high-energy particles move along Feynman paths,
which are practically rectilinear [18,19].} The solution of equation
$(2.8)$ can be found in the form

$$
F({\bf{r}},{\bf{r'}};s)=\frac{1}{(2\pi)^3}\int d {\bf{k}}
\exp{\biggl[W({\bf{r}};{\bf{k}};s)\biggl]}
e^{-i{\bf{k}}({\bf{r}}-{\bf{r'}})}. \eqno(3.1)
$$
Substituting $(3.1)$ into $(2.9)$ we have
$$
\exp {W (\bf{r}, \bf{k};s)}= 1 +
g\left\{\widehat{L_{r}}\left[V(\bf{r}; s)\exp{W(\bf{r},
\bf{k};s)}\right]+V(\bf{r};s)\exp{W(\bf{r},\bf{k};s)}K(\bf{k}^{2};s)\right\}.
\eqno (3.2)$$ Reducing this equation for the function $
W({\bf{r}};{\bf{k}};s)$, we get
$$
\exp{W({\bf{r}};{\bf{k}};s)}=1+g \widehat{L_{r}}\left\{V({\bf{r}},s)
\exp \left[W(\bf{r},\bf{k};s)-i\bf{k}\bf{r}\right]\right\}
e^{i{\bf{k}}{\bf{r}}}. \eqno (3.3)
$$
The function $W({\bf{r}};{\bf{k}};s)$ in exponent $ (3.1) $ can now
be written as an expansion in series in the coupling constant g:
$$
W({\bf{r}};{\bf{k}};s)= \sum_{n=1}^{\infty} g^{n}
W_{n}({\bf{r}};{\bf{k}};s). \eqno(3.4)
$$
Substituting $(3.4)$ into $(3.3)$ and using Taylor expansion, the
lhs.$(3.3)$ is rewritten as follow
$$1+\sum_{n=1}^\infty g^nW_n+\frac{1}{2!}\left(\sum_{n=1}^\infty
g^nW_n\right)^2+\frac{1}{3!}\left(\sum_{n=1}^\infty
g^nW_n\right)^3+\ldots,\eqno(3.5)$$ and the rhs.$(3.3)$ has form
$$
1+g\Biggl\{\hat{L}_r\left[V(\mathbf{r};s)\left(1+\sum_{n=1}^\infty
g^nW_n +\frac{1}{2!}\left(\sum_{n=1}^\infty
g^nW_n\right)^2+\frac{1}{3!}\left(\sum_{n=1}^\infty
g^nW_n\right)^3+\ldots\right)\right]+$$
$$
+V(\mathbf{r};s)\left[1+\sum_{n=1}^\infty g^nW_n
+\frac{1}{2!}\left(\sum_{n=1}^\infty
g^nW_n\right)^2+\frac{1}{3!}\left(\sum_{n=1}^\infty
g^nW_n\right)^3+\ldots\right]K(\mathbf{k};s)\Biggl\}.\eqno(3.6)
$$
From $(3.5)$ and $(3.6)$, to compare with two sides  of Eq.$ (3.3)$
following $ g $ coupling, we derive the following expressions for
the functions $W_{n}({\bf{r}};{\bf{k}};s)$
$$
W_{1}({\bf{r}};{\bf{k}};s)= \int d{\bf{q}}V({\bf{q}};s)
 K[({\bf{k}}+{\bf{q}})^{2};s] e^{-i{\bf{q}}{\bf{r}}}; \eqno(3.7)
$$
$$
W_{2}({\bf{r}};{\bf{k}};s)= -\frac{W_{1}^{2}({\bf{r}};{\bf{k}};s)}
{2!}+$$
$$
\frac{1}{2} \int d{\bf{q}}_{1} d{\bf{q}}_{2}
V({\bf{q}}_{1};s)V({\bf{q}}_{2};s) K[({\bf{k}}+ {\bf{q}}_{1}+
{\bf{q}}_{2})^2;s]\times
$$
$$\times [ K ({\bf{k}}+ {\bf{q}}_{1}; s)
+K[{\bf{k}}+{\bf{q}}_{2};s ]] e^{
-i{\bf{q}}_{1}{\bf{r}}-i{\bf{q}}_{2}\bf{r}}; \eqno(3.8)
$$
$$
W_{3}({\bf{r}};{\bf{k}};s)=
-\frac{W_{1}^{2}({\bf{r}};{\bf{k}};s)}{3!}+ \int
d{\bf{q}}_{1}d{\bf{q}}_{2}d{\bf{q}}_{3}
V({\bf{q}}_{1};s)V({\bf{q}}_{2};s)V({\bf{q}}_{3};s)
K[({\bf{k}}+{\bf{q}}_{1})^{2};s] \times$$
$$\times K[({\bf{k}}+
{\bf{q}}_{1}+{\bf{q}}_{2})^{2};s] K[({\bf{k}}+
{\bf{q}}_{1}+{\bf{q}}_{2}+{\bf{q}}_{3})^{2};s ]
e^{-i{\bf{q}}_{1}{\bf{r}}-i{\bf{q}}_{2}
{\bf{r}}-i{\bf{q}}_{3}{\bf{r}}}. \eqno(3.9)
$$

Oversleeves  by $ W_{1} $ only we obtain from Eqs $( 3.1)$, $(3.4)$
and $(2.3)$ the approximate expression for the scattering amplitude
$ [22] $

$$
T_{1}({\bf{p}},{\bf{p'}};s)=\frac{g}{(2\pi)^{3}} \int d{\bf{r}}
e^{i({\bf{p}}-{\bf{p'}}){\bf{r}}} V({\bf{r}},s)
e^{gW_{1}({\bf{r}},{\bf{p}},s)}. \eqno(3.10)
$$

To establish the meaning of this approximation, we expand $T_{1}$ in
a series in $g$:

$$
T_{1}^{(n+1)}({\bf{p}},{\bf{p'}};s)=\frac{g^{n+1}}{n!}\int
d{\bf{q}}_{1}...d{\bf{q}}_{n} V({\bf{q}}_{1};s)....V({\bf{q}}_{n};s)
$$
$$\times V({\bf{p}}-{\bf{p'}} -\sum_{i=1}^{n}
q_{i};s) \prod_{i=0}^{n} K[({\bf{q}}_{i}+{\bf{p'}})^{2};s].
\eqno(3.11)
$$

Let us compare Eq. $(3.10)$ with the $ (n+1)-th $ iteration term of
exact Eq. $(2.1)$

$$
T^{(n+1)}({\bf{p}},{\bf{p'}};s)=\int
d{\bf{q}}_{1}...d{\bf{q}}_{n}V({\bf{q}}_{1};s)... V({\bf{q}}_{n};s)
V({\bf{p}}-{\bf{p'}} -\sum_{i=1}^{n} q_{i};s)\times
$$
$$ \sum_{p}K[({\bf{q}}_{1}+{\bf{p'}})^{2};s]
K[({\bf{q}}_{1}+ {\bf{q}}_{2}+{\bf{p'}})^{2};s]... K[(
\sum_{i=1}{\bf{q}}_{i}+{\bf{p'}})^{2};s], \eqno(3.12)
$$
where $\sum_{p}$ is the sum over the permutations of the momenta
${\bf{p}}_{1}$ ,${\bf{p}}_{2}...$ ${\bf{p}}_{n}$. It is readily seen
from $(3.11)$ and $(3.12)$ that our approximation in the case of the
Lippmann-Schwinger equation is identical with the
${{\bf{q}}_{i}}{{\bf{q}}_{j}}$ approximation.\\

\section{ Asymptotic behavior of the scattering amplitude
at high energies}

In this section  the solution of the Logunov-Tavkhelidze
quasi-potential equation obtained in the previous section for the
scattering amplitude can be used to find the asymptotic behavior as
$ s\rightarrow \infty $ for fixed $ t $. In the asymptotic
expressions we shall retain both the principal term and the
following term, using the formula

$$
e^{W({\bf{r}},{\bf{p'}}; s)}= e^{W_{1}({\bf{r}},{\bf{p'}};
s)}\biggl [1+g^{2}W_{2}({\bf{r}},{\bf{p'}};s)+...\biggl],
\eqno(4.1)
$$
where $ W_{1}$ and $ W_{2}$ are given by $(3.7)$ and $(3.8)$.\\
We take the $ z $ axis along the vector  $ ({\bf{p}}+{\bf{p'}}) $
then

$$
{\bf{p}}-{\bf{p'}}={\bf{{\Delta}_{\perp}}};\quad
{\bf{\Delta}_{\perp}} {\bf {n}_{z}}=0;\quad t=-
{\bf{\Delta}_{\perp}^{2}}. \eqno(4.2)
$$

Noting

$$
K({\bf{p}}+{\bf{p'}};s) =\frac{1}
{\sqrt{({\bf{p}}+{\bf{p'}})^{2}+m^{2}}}\frac{1}
{({\bf{p}}+{\bf{p'}})^{2}-\frac{s}{4}+m^{2}-i\varepsilon}\Biggr
|_{s\rightarrow\infty;t- fixed} =$$
$$
=\frac{2}{s(q_{z}^{2}-i\varepsilon)} \left[1-\frac{ 3q_{z}^{2}+
{\bf{q}_{\perp}}^{2} + {\bf{q}_{\perp}}{\bf{\triangle}_{\perp}}}
{\sqrt{s}(q_{z}-i\epsilon)}\right] +
 O\left(\frac{1}{s^{2}}\right), \eqno(4.3)
$$

and using  Eqs $(3.4)$ $(3.7)$ and $(3.8)$ we obtain

$$
W_{1}=\biggl (\frac{W_{10}}{s}\biggl ) + \biggl
(\frac{W_{11}}{s\sqrt{s}}\biggl ) + O \biggl(\frac{1}{s^{2}}\biggl
); \eqno(4.4)
$$

$$
W_{2}=\biggl (\frac{W_{20}}{s^{2}\sqrt{s}}\biggl ) +O \biggl
(\frac{1}{s^{3}}\biggl ), \eqno(4.5)
$$
where

$$
W_{10}=2\int d{\bf{q}} V({\bf{q}};s) \frac{e^{i{\bf{q}}{\bf{r}}}}
{(q_{z}^{2}-i\varepsilon)^{2}} = 2i\int_{-\infty}^{z} dz'
V(\sqrt{{\bf{q}_{\perp}}^{2}+z'^{2}};s); \eqno (4.6) $$

$$
W_{11}=-2 \int d{\bf{q}}V({\bf{q}};s) e^{-i{\bf{q}}{\bf{r}}} \frac{
3q_{z}^{2}+ {\bf{q}_{\perp}}^{2} +
{\bf{q}_{\perp}}{\bf{\triangle}_{\perp}}} {(q_{z}-i\epsilon)^2} =$$
$$ =-6 V(\sqrt{{\bf{q}_{\perp}}^{2}+z'^{2}};s) + 2(
-{\bf{{\nabla}_{\perp}}}^{2}-i\bf{q}_{\perp}{\bf{{\nabla}_{\perp}}})
\int_{-\infty}^{z} dz' V(\sqrt{{\bf{q}_{\perp}}^{2}+z'^{2}};s);
\eqno(4.7)$$

$$
W_{20}=-4 \int d{\bf{q}}_{1}d{\bf{q}}_{2}
e^{-i({\bf{q}}_{1}+{\bf{q}}_{2}){\bf{r}}}
V({\bf{q}}_{1};s)V({\bf{q}}_{2};s)
\frac{3q_{1z}q_{2z}+{\bf{q}}_{1\perp}{\bf{q}}_{2\perp}}
{(q_{1z}-i\varepsilon)(q_{2z}-i\varepsilon)(q_{1z}+q_{2z}-i\varepsilon)}$$
$$= -4i\left\{ 3 \int_{-\infty}^{z} dz'
V^{2}(\sqrt{{\bf{q}_{\perp}}^{2}+z'^{2}};s)+\left[
{\bf{{\nabla}_{\perp}}}\int_{-\infty}^{z'} dz''
V^2(\sqrt{{\bf{q}_{\perp}}^{2}+z''^{2}};s)\right]^{2}\right\}.
\eqno(4.8)
$$

In the limit $s\rightarrow \infty $ and $ (t/s) \rightarrow 0 $ $
W_{10}$ makes the main contribution, and the remaining terms are
corrections. Therefore, the function $ \exp {W}$ can be represented
by means of the expansion $(4.1)$ where $W_{10}$, $ W_{11}$ and $
W_{20}$ are determined by Eqs. $(4.6)-(4.8)$ respectively. The
asymptotic behavior scattering amplitude can be written in the
following form

$$
T(\bf{p},\bf{p'}; s)=\frac{g}{(2\pi)^{3}}\int d^{2}\bf{r}_{\perp}dz
e^{i{\bf{\Delta_{\perp}}}{\bf{r_{\perp}}}}V(\sqrt{\bf{r}^{2}+z^{2}};s)\times$$
$$\times \exp\biggl(g\frac{W_{10}}{s}\biggl)
\biggl(1+g\frac{W_{11}}{s\sqrt{s}}+g^{2}\frac{W_{20}}{s^{2}\sqrt{s}}+...\biggl).
\eqno(4.9)
$$

Substituting $(4.6)-(4.8)$ into $(4.9)$ and making calculations, at
high energy $s\rightarrow\infty$ and fixed momentum transfers
$(t/s)\rightarrow 0 $, we finally obtain[22]

$$
T(s,t)=\frac{g}{2i(2\pi)^{3}} \int
d^{2}\bf{r}_{\perp}e^{i{\bf{\Delta_{\perp}}}{\bf{r_{\perp}}}}
\Biggl\{e^{\bigl[\frac{2ig}{s}\int_{-\infty}^{\infty}dz
V(\sqrt{\bf{r}^{2}+z^{2}};s)\bigl]}-1 \Biggl\}-
$$

$$
-\frac{6g^{2}}{(2\pi)^{3}s\sqrt{s}} \int d^{2}\bf{r}_{\perp}
e^{i{\bf{\Delta_{\perp}}}{\bf{r_{\perp}}}} \exp\biggl[\frac{2ig}{s}
\int_{-\infty}^{\infty} dz'
V(\sqrt{\bf{r}_{\perp}^{2}+z^{2}};s)\biggl] \int_{-\infty}^{\infty}
dz V(\sqrt{\bf{r}_{\perp}^{2}+z^{2}};s)-$$
$$
-\frac{ig}{(2\pi)^{3}\sqrt{s}}\int d^{2}\bf{r}_{\perp}
e^{i{\bf{\Delta_{\perp}}}{\bf{r_{\perp}}}} \times$$
$$\times\int_{-\infty}^{\infty}dz \Biggl \{\exp \biggl[\frac{2ig}{s}
\int_{z}^{\infty} dz'V(\sqrt{\bf{r}_{\perp}^{2}+z'^{2}};s) \biggl] -
\exp{\biggl[\frac{2ig}{s} \int_{-\infty}^{\infty} dz'
V(\sqrt{\bf{r}_{\perp}^{2}+z'^{2}};s)\biggl]}\Biggl\}\times$$
$$\times \Biggl\{\int_{z}^{\infty} dz'
{\bf{\nabla_{\perp}}}^{2}V(\sqrt{\bf{r}_{\perp}^{2}+z'^{2}};s)-
\frac{2ig}{s}
\biggl[\int_{z}^{\infty}dz'{\bf{\nabla_{\perp}}}V(\sqrt{\bf{r}_{\perp}^{2}+z^{2}};s)
\biggl]^{2}\Biggl\}$$
$$-\frac{2ig}{(2\pi)^{3}s}{\bf{\Delta_{\perp}}}^{2}
\int d^{2}\bf{r}_{\perp}V(\sqrt{\bf{r}_{\perp}^{2}+z'^{2}};s)]
e^{i{\bf{\Delta_{\perp}}}{\bf{r_{\perp}}}}\times$$
$$
\times\int_{-\infty}^{\infty} zdz
V(\sqrt{\bf{r}_{\perp}^{2}+z^{2}};s) \exp {\biggl[ \frac{2ig}{s}
\int_{z}^{\infty} dz'V(\sqrt{\bf{r}_{\perp}^{2}+z'^{2}};s)\biggl ]}.
\eqno(4.10)
$$

In this expression $(4.10)$ the first term describes the leading
eikonal behavior of the scattering amplitude, while the remaining
terms determine the corrections of relative magnitude $1/\sqrt{s}$.
The similar result Eq.$(4.10)$ is also found  by means of the
functional integration $[20]$.\\

As is well known from the investigation of the scattering amplitude
in the Feynman diagrammatic technique, the high-energy asymptotic
behavior can contain only logarithms and integral powers of $ s $. A
similar effect is observed here, since integration of the expression
$(4.10)$  leads to the vanishing of the coefficients for
half-integral powers of $s$. Nevertheless, allowance for the terms
that contain the half-integral powers of $s$ is needed for the
calculations of the next corrections in the scattering amplitude,
and leads to the appearance of the so-called retardation effects,
which are absent in the  principal asymptotic term.\\

    In the limit of high energies $s\rightarrow \infty $ and
for fixed momentum transfers $t$ the expression for the scattering
amplitude within the framework of the functional - integration
method takes the Glauber form with eikonal function corresponding to
a Yukawa interaction potential between "nucleons". Therefore,  the
local quasi-potential for the interaction between the "nucleons"
from perturbation theory in that region can be chosen by following
forms. For the scalar meson exchange the quasi-potential decreases
with energy

$$
V(r;s)=-\frac{g^2}{8\pi s}\frac{e^{-\mu r}}{r}. \eqno(4.11)
$$

The first term  in the expression $(4.10)$ describes the leading
eikonal behavior of the scattering amplitude. Using integrals
calculated in the Appendix, we find

$$
T^{(0)}_{Scalar}(s,t)=-\frac{g}{2i(2\pi)^3}\int d^2\mathbf{r}_\bot
e^{i\Delta_\bot \mathbf{r}_\bot}
\left\{\exp\left[\frac{2ig}{s}\int_{-\infty}^{+\infty} dz
V\left(\sqrt{\mathbf{r}_\bot^2+z^2};s\right)\right]-1\right\} $$
$$
=\frac{g^4}{4(2\pi)^4s^2}\left[\frac{1}{\mu^2-t}-\frac{g^3}{8(2\pi)^2
s^2}F_{1}(t)+\frac{g^6}{48(2\pi)^5s^4}F_2(t)\right]. \eqno(4.12)$$

The next term in $(4.10)$ describes first correction to the leading
eikonal amplitude

$$
T^{(1)}_{Scalar}(s,t)=-\frac{6g^2}{(2\pi)^3s\sqrt{s}}\int d^2r_\bot
e^{i\Delta_\bot r_\bot}
\exp\left[\frac{2ig}{s}\int_{-\infty}^{+\infty} dz
V\left(\sqrt{\mathbf{r}_\bot^2+z^2};s\right)\right]\times $$
$$
\times\int_{-\infty}^{+\infty} dz
V\left(\sqrt{\mathbf{r}_\bot^2+z^2};s\right)=$$
$$=\frac{3g^4}{4(2\pi)^6s^2\sqrt{s}}\left[\frac{2}{\mu^2-t}
-\frac{g^3}{2(2\pi)^2s^2}F_{1}(t)+\frac{g^6}{8(2\pi)^5s^4}F_{2}(t)\right]\eqno(4.13)
$$
where
$$
F_1(t)= \frac{1}{t\sqrt{1-\frac{4\mu^2}{t}}}ln\left|
\frac{1-\sqrt{1-4\mu^2/t}}{1+\sqrt{1-4\mu^2/t}}\right|\eqno(4.14)
$$
and

$$
F_2(t)=\int_0^1
dy\frac{1}{(ty+\mu^2)(y-1)}ln\left|\frac{\mu^2}{y(ty+\mu^2-t)}\right|\eqno(4.15)
$$

A similar calculations can be applied  for other exchanges with
different spins. In the case of the vector model $ L_{int}=
-g\varphi^{\star}i\partial_\sigma\varphi
A^{\sigma}+g^{2}A_{\sigma\sigma}A^{\sigma}\varphi\varphi^{\star}\varphi$
the quasi-potential is independent of energy $V(r; s)=
-(g^2/4\pi)(e^{-\mu r}/ r)$, we find

$$
T^{(0)}_{Vector}(s,t)=\frac{g^4}{2(2\pi)^4s}\left[\frac{1}{\mu^2-t}-\frac{g^3}{4(2\pi)^2
s}F_{1}(t)+\frac{g^6}{12(2\pi)^5s^2}F_2(t)\right]\eqno(4.16)
$$

$$
T^{(1)}_{Vector}(s,t)=\frac{3g^4}{2(2\pi)^6s\sqrt{s}}\left[\frac{2}{\mu^2-t}
      -\frac{g^3}{(2\pi)^2s}F_{1}(t)+\frac{g^6}{2(2\pi)^5s^2}F_{2}(t)\right]\eqno(4.17)
$$

In the case of tensor model \footnote{The model of interaction of a
scalar "nucleons" field $\varphi(x) $ with a gravitational field $
g_{\mu\nu}(x)$ in the linear approximation to $
h^{\mu\nu}(x)$;$[18]$ $ L(x)=L_{0,{\varphi}}(x)+L_{0,grav.}(x)
+L_{int}(x) $,where

$$ L_{0}(x)=\frac{1}{2} [\partial^{\mu} \varphi (x)\partial_{\mu} \varphi(x)
-m^{2} {\varphi}^{2}(x) ] ,$$

$$ L_{int}(x)=-\frac{\kappa}{2} h^{\mu\nu}(x)T_{\mu\nu}(x) ,$$

$$ T_{\mu\nu}(x)=\partial_{\mu}\varphi(x) \partial_{\nu}\varphi(x)-
\frac{1}{2}\eta_{\mu\nu} [\partial^{\sigma} \varphi
(x)\partial_{\sigma} \varphi(x) -m^{2} {\varphi}^{2}(x) ] ,$$

$ T_{\mu\nu}(x)$-the energy momentum tensor of the scalar field. The
coupling constant $\kappa$ is related to Newton's constant of
gravitation $ G $ by $ \kappa^2=16 \pi G $ }, the quasi- potential
increases with energy $V(r; s)= (\kappa^2 s/2\pi)(e^{-\mu r}/r)$, we
have

$$
T^{(0)}_{Tensor}(s,t)=-\frac{\kappa^4}{(2\pi)^4}\left[\frac{1}{\mu^2-t}
  +\frac{\kappa^3}{2(2\pi)^2}F_{1}(t)+\frac{\kappa^6}{3(2\pi)^5}F_2(t)\right]\eqno(4.18)
$$
$$
T^{(1)}_{Tensor}(s,t)=-\frac{3\kappa^4}{(2\pi)^6\sqrt{s}}\left[\frac{2}{\mu^2-t}
      +\frac{2\kappa^3}{(2\pi)^2}F_{1}(t)+\frac{2\kappa^6}{(2\pi)^5}F_{2}(t)\right]\eqno(4.19)
$$

 To conclude this section it is important to note that
in the framework of standard field theory for the high-energy
scattering, different methods have been developed to investigate the
asymptotic behavior of individual Feynman diagrams and their
subsequent summation. In different theories including quantum
gravity the calculations of Feynman diagrams in the eikonal
approximation is proceed  in a similar way as analogous the
calculations in QED. Reliability of the eikonal approximation
depends on spin of the exchanges field $[5,6]$. The eikonal captures
the leading behavior of each order in perturbation theory, but the
sum of leading terms is subdominant to the terms neglected by this
approximation. The reliability of the eikonal amplitude for gravity
is uncertain [14]. Instead of the diagram technique perturbation
theory, our approach is based on the exact expression of the
scattering amplitude and modified perturbation theory which in
lowest order contains the leading eikonal amplitude and the next
orders are its corrections.

\section{Relationship between the operator and Feynman path methods }

What actual physical picture may correspond to our result given by
Eq. $(4.10)$ ? To answer  this question we establish the
relationship between the operator and Feynman path methods in Ref.
[31], which treats the quasi-potential equation in the language of
functional integrals. The solution of this equation can be written
in the symbolic form:

$$
\exp (W)=\frac{1}{1-gK[(-i {\bf{{\nabla}}}-{\bf{k}})^{2}] V({\bf
{r}})}\times \bf{1}=
$$
$$ =-i\int_{0}^{\infty}d\tau \exp [i\tau(1+i\varepsilon)]
\exp \left\{-i\tau g K[(-i {\bf{{\nabla}}}-{\bf{k}})^{2}]
V({\bf{r}})\right\}\times \bf{1}. \eqno(5.1) $$

In accordance with the Feynman parametrization $[31]$, we introduce
an ordering index $ \eta $ and write Eq. $(5.1)$ in the form

$$
\exp (W)= -i\int_{0}^{\infty} d\tau e^{i\tau
(1+i\varepsilon)}\exp\left\{ -ig \int_{0}^{\infty} d\eta
K[(-i{\bf{{\nabla}_{\eta+\varepsilon}}}-{\bf{k}})^{2}]
U({\bf{r}}_{\eta})\right\} \times \bf{1}. \eqno(5.2)
$$

Using Feynman transformation

$$
F[P(\eta)]= \int D{\bf{p}}\int_{x(0)=0} \frac{D{\bf{x}}}{(2\pi)^{3}}
\exp\left\{i \int_{0}^{\tau} d\eta \dot{{\bf{r}}}(\eta)[
{\bf{p}}(\eta)-P(\eta)]\right\} F[{\bf{p}}(\eta)], \eqno(5.3)
$$

we write the solution of  Eq. $(2.8)$ in the form of the functional
integral

$$
\exp (W)=-i\int_{0}^{\infty} d\tau e^{i\tau
(1+i\varepsilon)}\times$$
$$ \times \int D{\bf{p}}\int_{x(0)=0}
\frac{D{\bf{x}}}{(2\pi)^{3}} \exp\left\{i \int_{0}^{\tau} d\eta
\dot{{\bf{x}}}(\eta)[ {\bf{p}}(\eta)-P(\eta)]\right\}
G({\bf{x}},{\bf{p}};\tau)\times \bf{1}.\eqno(5.4)
$$

In Eq.$(5.4)$ we enter the function $G $

$$
G({\bf{x}},{\bf{p}};\tau)=\exp \left\{-i \int_{0}^{\tau}d\eta
\dot{{\bf{x}}}(\eta) \nabla_{\eta+\varepsilon}\right\}\times
\exp\left\{ -ig \int_{0}^{\tau} d\eta
K[({\bf{p}}(\eta)-{\bf{k}})^{2}]V({\bf{r}}_{\eta})\right\}, \eqno
(5.5)
$$

which satisfies the equation

$$
\frac{dG}{d\tau}= \left\{ -igK
[({\bf{p}}(\tau)-{\bf{k}})^{2}]V({\bf{r}}-
\dot{{\bf{x}}}(\tau-\varepsilon)){\bf{{\nabla}}}\right\}G;
$$
$$
G(\tau=0)=1. \eqno(5.6)
$$

Finding from  Eq. $(5.6)$  the operator function $ G $ and
substituting it into Eq. $(5.6)$ for $ W $ we obtained the following
final expression:

$$
\exp (W)=-i\int_{0}^{\infty} d\tau e^{i\tau (1+i\varepsilon)}\int D
{\bf{p}} \frac{1}{(2\pi)^{3}}\int_{x(0)=0}
\frac{D{\bf{x}}}{(2\pi)^{3}} \exp\left\{i \int_{0}^{\tau} d\eta
\dot{{\bf{x}}}(\eta)p(\eta)\right\} \exp (g \prod), \eqno (5.7)
$$

where

$$
\prod =-i\int_{0}^{\infty} d\tau K [({\bf{p}}(\eta)-{\bf{k}})^{2}]
V\left[{\bf{r}}-\int_{0}^{\tau} d\xi \dot{{\bf{x}}(\xi)}\vartheta
(\xi-\eta+\varepsilon)\right]; \eqno (5.8)
$$

$$
{\prod}^{2}=-\int_{0}^{\tau_{1}}\int_{0}^{\tau_{2}}d\tau_{1}d\tau_{2}
K [({\bf{p}}(\eta_{1})-{\bf{k}})^{2}]K
[({\bf{p}}(\eta_{2})-{\bf(k)})^{2}]\times$$
$$
\times V\left[{\bf{r}_{1}}-\int_{0}^{\tau_{1}} d\xi
\dot{{\bf{x}}(\xi)}\vartheta (\xi-\eta+\varepsilon)\right]
V\left[{\bf{r}_{2}}-\int_{0}^{\tau_{2}} d\xi
\dot{{\bf{x}}(\xi)}\vartheta
(\xi-\eta+\varepsilon)\right].\eqno(5.9)
$$

Writing out the expansion $[2,3]$

$$
\exp(W)=\overline{\exp(g\prod)}= \exp({g\overline{\prod}})
\sum_{n=0}^{\infty}\frac{g^{n}}{n!}\overline{\Bigl(\prod-\overline{\prod}\Bigl)^{n}},\eqno
(5.10)
$$
in which the sign of averaging denoted integration with respect to $
\tau $, $ \bf{x}(\eta)$ and $ \bf{p}(\eta)$ with the corresponding
measure ( see, for example Eq. $(5.7)$ ), and performing the
calculations, we find

$$
W_{1}=\overline{ \prod }, \eqno (5.10a)
$$

$$
W_{2}=\frac{\overline{{\prod}^{2}}- {\overline{ \prod }}^{2} }
{2!},\eqno (5.10b)
$$

$$
W_{3}=\frac{1}{3!}\Bigl[
\overline{{\prod}^{3}}-{\overline{\prod}}^{3}-
3\overline{\prod}(\overline{{\prod}^{2}}-{\overline{\prod}}^{2})\Bigl],
etc. \eqno(5.10c)
$$
i.e. the expressions $(5.10)$ and $(4.1)$ are identical.
$$
W_{1}=\overline{ \prod }=-i\int_{0}^{\infty} d\tau K
[({\bf{p}}(\eta)-{\bf{k}})^{2}] \exp \left[-\int_{0}^{\tau} d\xi
\dot{{\bf{x}}(\xi)}\vartheta(\xi-\eta+\varepsilon)\nabla_{\eta}\right]V(\overrightarrow{r})
$$
$$
=\int d{\bf{q}} e^{-{\bf{q}}{\bf{r}}} K[({\bf{q}}+{\bf {k}})^{2}]
V({\bf{q}};s); \eqno (5.11)$$

$$ \overline{{\prod}^{2}}=K[({\bf{\nabla}_{\bf{r_{1}}}}+
\bf{\nabla}_{\bf{r_{2}}}+\bf{k})^{2}]
K[(\bf{\nabla}_{\bf{r_{1}}}+\bf{k})^{2}]
K[(\bf{\nabla}_{\bf{r_{2}}}+\bf{k})^{2}] V(\bf{r_{1}};s)
V(\bf{r_{2}};s)
$$
$$ =\int d {\bf{q_{1}}}\int d {\bf{q_{2}}}
e^{-i({\bf{q_{1}}}+{\bf{q_{2}}}){\bf{r}}}
K[({\bf{q_{1}}}+{\bf{q_{2}}}+{\bf{k}})^{2}]
\left\{K[({\bf{q_{1}}}+{\bf{k}})^{2}]+
K[({\bf{q_{2}}+{\bf{k}}})^{2}]\right\} V({\bf{r_{1}}};s)
V({\bf{r_{2}}};s); \eqno(5.12)
$$
$$
W_{2}=-\frac{W_{1}^{2}}{2!}+ \frac{1}{2}\int
d{\bf{q_{1}}}d{\bf{q_{2}}} V({\bf{q_{1}}})V({\bf{q_{2}}})
\{K[({\bf{q_{1}}}+{\bf{k}})^{2};s]+
K[({\bf{q_{2}}}+{\bf{k}})^{2};s]\}; \eqno(5.13)
$$

$$
W_{3}=-\frac{W_{1}^{3}}{3!} +  \int
d{\bf{q}}_{1}d{\bf{q}}_{2}d{\bf{q}}_{3}
V({\bf{q}}_{1};s)V({\bf{q}}_{2};s)V({\bf{q}}_{3};s)
K[({\bf{k}}+{\bf{q}}_{1})^{2};s] \times$$
$$\times K[({\bf{k}}+
{\bf{q}}_{1}+{\bf{q}}_{2})^{2};s] K[({\bf{k}}+
{\bf{q}}_{1}+{\bf{q}}_{2}+{\bf{q}}_{3})^{2};s ]
e^{-i{\bf{q}}_{1}{\bf{r}}-i{\bf{q}}_{2}
{\bf{r}}-i{\bf{q}}_{3}{\bf{r}}}; etc \eqno(5.14)
$$

Restricting ourselves in the expansion $(5.10)$ to the first term
$(n=0) $, we obtain the approximate expression $(4.12)$ for the
scattering amplitude, which corresponds to the allowance for the
particle Feynman paths. These paths can be considered as a classical
paths and coincide in the case of the scattering of high-energy
particles through small angles to straight-line paths trajectories.

\section{Conclusions}

Asymptotic behavior of scattering amplitude for  two scalar
particles  at high energy and fixed momentum transfers was studied.
In the framework of quasi-potential approach and the modified
perturbation theory the systematic scheme of finding the corrections
to the principal asymptotic leading scattering amplitudes was
constructed and developed. Results obtained by two different
approaches (quasi-potential and functional) for this problem, as it
has shown that they are identical. Results obtained by us are
extended to the case of scalar particles of the field $\varphi(x)$
interacting with a  vector and gravitational fields. The first
correction to the leading eikonal scattering amplitude in quantum
field theory was obtained.

\section{Acknowledgements}

We are  grateful to Profs. B.M.Barbashov,
V.V.Nesterenko,V.N.Pervushin for valuable discussions and Prof.
G.Veneziano for suggesting this problem and his encouragement.
N.S.H. is also indebted to Profs. I.T. Todorov and H. Fried for
reading the manuscript and making useful remarks for improvements.
This work was supported in part by the International Center for
Theoretical Physics, Trieste, the Abdus Salam International Atomic
Energy Agency, the United Nations Educational, Scientific and
Cultural Organization, and the Vietnam National Research Programme
in Natural Science No 406406.

\vskip0.5cm \noindent {\Large \bf Appendix A: The kernel of the
quasi-potential equation $[25]$}\vskip0.5cm

 We denote by $G(\bf{p},\bf{p'},\varepsilon_p,\varepsilon_q,E)$ the total Green
 function for two particles, where $\bf{p}$ and $\bf{p'}$ are
 the momenta of the initial and final states in c.m.s and $2E=\sqrt{s}$ is
 the total energy (see fig.1).

\begin{center}
\begin{picture} (200,30)(-100,0)
\GCirc(0,0){20}{0.6} \ArrowLine(-35,35)(-14.14,14.14)
\ArrowLine(14.14,-14.14)(35,-35) \ArrowLine(-35,-35)(-14.14,-14.14)
\ArrowLine(14.14,14.14)(35,35)
\Text(-60,40)[]{$E+\varepsilon_p,\vec{p}$}
\Text(60,40)[]{$E+\varepsilon_{p'},\vec{p'}$}
\Text(-60,-40)[]{$E-\varepsilon_p,-\vec{p}$}
\Text(60,-40)[]{$E-\varepsilon_{p'},-\vec{p'}$}
\end{picture}\vspace{2cm}\\
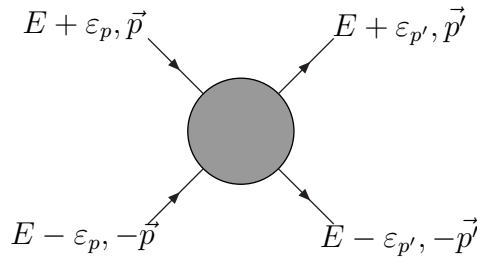
{\textbf{Fig.1}}. \vspace{0.2cm}
\end{center}

In these notations the Bethe-Salpeter equation is of the form

$$
G(\bf{p},\bf{p'},\varepsilon_p,\varepsilon_{p'},E)=iF(\bf{p},\varepsilon_p,E)
\delta(\bf{p}-\bf{p'})\delta(\varepsilon_p-\varepsilon_{p'})$$

$$+F(\bf{p},\varepsilon_p, E)\int K(\bf{p},\bf{q},\varepsilon_p,\varepsilon, E)
G(\bf{q},\bf{p'},\varepsilon,\varepsilon_{p'}, E) d\bf{q}
d\varepsilon,\eqno(A.1)$$ where
$$iF(\bf{p},\varepsilon_p,
E)=\frac{2}{\pi}D(E+\varepsilon_p,\bf{p})D(E-\varepsilon_p,\bf{p})$$
$$D(E+\varepsilon_p,\bf{p})=\frac{1}{(E+\varepsilon_p)^2-p^2-m^2+i\epsilon}.\eqno(A.2)$$
Now we introduce formally the scattering amplitude $T$ which on the
mass - shell $\varepsilon_p=\varepsilon_{p'}=0, \quad
p^2=p'^2=E^2-m^2$ gives the physical scattering amplitude:

$$G(\bf{p},\bf{p'},\varepsilon_p,\varepsilon_{p'}, E)-iF(\bf{p},\varepsilon_p, E)
\delta(\bf{p}-\bf{p'})\delta(\varepsilon_p-\varepsilon_{p'})= $$
$$=iF(\bf{p},\varepsilon_p,E)T(\bf{p},\bf{p'},\varepsilon_p,\varepsilon_{p'},E)
F(\bf{p'},\varepsilon_{p'},E).\eqno(A.3)$$

Then inserting $(A.3)$ into $(A.1)$, we get for $T$ the equation

$$
T(\bf{p},\bf{p'},\varepsilon_p,\varepsilon_{p'}, E) =
K(\bf{p},\bf{p'},\varepsilon_p,\varepsilon_{p'}, E)$$
$$ +\int d\bf{q} d\varepsilon K(\bf{p},\bf{q},\varepsilon_p,\varepsilon, E)
F(\bf{q},\varepsilon,E)
T(\bf{q},\bf{p'},\varepsilon,\varepsilon_{p'},E).\eqno(A.4)$$

We wish to obtain an equation of the Lippmann - Schwinger type for a
certain function $T(\bf{p},\bf{p'},E)$ which on the mass-shell
$p^2=p'^2=E^2-m^2$ would give the physical scattering amplitude:

$$
T(\bf{p},\bf{p'},E)=V(\bf{p},\bf{p'},E)+\int
d\bf{q}V(\bf{p},\bf{q},E)F(\bf{q},E)T(\bf{q},\bf{p'},E),\eqno(A.5)
$$
where
$$
F(\bf{q},E)=\int d\varepsilon
F(\bf{q},\varepsilon,E)=-\frac{2i}{\pi}\int d\varepsilon
\frac{1}{(E+\varepsilon)^2-p^2-m^2+i\epsilon}\times
 \frac{1}{(E-\varepsilon)^2-p^2-m^2+i\epsilon}$$
 $$=\frac{1}{\sqrt{q^2+m^2}(q^2+m^2-E^2)}.\eqno(A.6)
$$
On the mass-shell, the total energy $E=\frac{\sqrt{s}}{2}$, we
receive the kenel that is brought out in Eq.(2.1)
$$K(q^2;s)\equiv F\left(q,E=\frac{\sqrt{s}}{2}\right)
=\frac{1}{\sqrt{q^2+m^2}(q^2+m^2-\frac{s}{4})}. \eqno{(A.7)}$$
  This can be achieved by a
conventional choice of the potential $ V(\bf{p},\bf{p'},E)$, which
can obviously be made by different methods.There are two methods
that have been suggested for constructing a complex potential
dependent on energy with the help of which one can obtain from an
equation of the Schr\"{o}dinger
type the exact scattering amplitude on the mass - shell. \\
The first method is based on the two-time Green function $ [23]$
which in the momentum space is defined

$$ G(\bf{p},\bf{p'},E)=\int d\varepsilon_p d\varepsilon_{p'}
G(\bf{p},\bf{p'},\varepsilon_p,\varepsilon_{p'},E).\eqno(A.8)$$

Then using $(A.3)$ and $(A.8)$ we can determine the corresponding
off-shell scattering amplitude

$$T_1(\bf{p},\bf{p'},E)=\frac{1}{F(\bf{p},E)F(\bf{p'},E)}\int
F(\bf{p},\varepsilon_p,E)
T(\bf{p},\bf{p'},\varepsilon_p,\varepsilon_{p'},E) F(\bf{p'},
\varepsilon_{p'},E) d\varepsilon_p d\varepsilon_{p'}.\eqno(A.9)$$

From expression $(A.9)$ it is directly seen that $T$ on mass-shell
$p^2=p'^2=E^2-m^2 $ coincides with the scattering amplitude $
T(\bf{p},\bf{p'},0,0,E)\equiv T(\bf{p},\bf{p'},E)$. The potential
$V_1$ for Eq.$(A.5)$ is constructed by iteration of Eqs.$(A.4)$,
$(A.5)$ and $(A.9)$. In particular, in the lowest order, we have

$$
V_1(\bf{p},\bf{p'},E)=\frac{1}{F(\bf{p},E)F(\bf{p'},E)}
 \int F(\bf{p},\varepsilon_p,E)K(\bf{p},\bf{p'},\varepsilon_p,\varepsilon_{p'},E)
 F(\bf{p'}, \varepsilon_{p'},E) d\varepsilon_p
 d\varepsilon_{p'}.\eqno(A.10)
 $$

The second method consists in constructing the potential $V_2$ for
Eq.$(A.4)$ by means of the scattering amplitude T on the mass-shell
obtained by perturbation theory, e.g. from Eq.$(A.4)$ and the
iterations of Eq.$(A.5)$ accompanied by the transition to the
mass-shell.\\

We write down Eq.$(A.5)$ in the symbolic form $ T_2=V_2+V_2\times
T_2$ and obtain in the lowest orders of $V_2$ the expressions

$$V_2^{(2)}=[T^{(2)}], V^{(4)}=[T^{(4)}]-[V_2^{(2)}\times T_2^{(2)}],$$
$$V_2^{(6)}=[T^{(6)}]-[V_2^{(2)}\times T_2^{(4)}]-[V_2^{(4)}\times
T_2^{(2)}]\ldots,\eqno(A.11)$$ where the square brackets mean here
the transition to the mass-shell. Hence, it follows that in the
second method we get a local potential dependent only on
$(\bf{p}-\bf{p'})^2$ and $E$ and in r-space on r and $E$.\\

We shall consider, as an example, the application of the above
methods to a model of quantum field theory, in which scalar
particles of mass m interact by exchanging scalar "photons" of small
mass $\mu$. We shall put $\mu=0$ where it is possible. In the ladder
approximation (without considering the crossing symmetry) the kernel
of Eq.$(A.5)$ is of the form

$$K(\bf{p},\bf{p'},\varepsilon_p,\varepsilon_{p'},E)
 = -\frac{ig^2}{(2\pi)^4}\frac{1}{(\varepsilon_p-\varepsilon_{p'})^2
 -(\bf{p}-\bf{p'})^2-\mu^2+i\epsilon}.\eqno(A.12)$$

\vskip0.5cm \noindent {\Large \bf Appendix B: Some integrals used in
this paper} \vskip0.5cm

Firstly, we consider the integral

$$
I_1=\int_{-\infty}^\infty dz
V\left(\sqrt{\mathbf{r}_\bot^2+z^2};s\right)= -\frac{g^2}{8\pi
s}\int_{-\infty}^\infty dz\frac{e^{-\mu
\left(\sqrt{\mathbf{r}_\bot^2+z^2}\right)}}
{\sqrt{\mathbf{r}_\bot^2+z^2}} \eqno(B.1)
$$

we have

$$
I_1=-\frac{g^2}{4(2\pi)^4 s}\int d^3p\int_{-\infty}^{+\infty}dz
\frac{e^{i\textbf{pr}}}{\mu^2+p^2}= -\frac{g^2}{4(2\pi)^4 s}\int
d^3p\int_{-\infty}^{+\infty}dz \frac{e^{i(p_\bot r_\bot +
p_{//}z)}}{\mu^2+p^2}=
$$
$$
=-\frac{g^2}{4(2\pi)^4 s}\int d^3p \frac{e^{i(p_\bot r_\bot
)}}{\mu^2+p^2}\int_{-\infty}^{+\infty}dz e^{ip_{//}z}=
$$
$$
=-\frac{g^2}{4(2\pi)^4 s}\int d^2p_\bot dp_{//}\frac{e^{i(p_\bot
r_\bot )}}{\mu^2+p_\bot^2+p_{//}^2}\times(2\pi)\delta(p_{//})=
$$
$$
=-\frac{g^2}{4(2\pi)^3 s}\int d^2p_\bot e^{i(p_\bot r_\bot )}\int
dp_{//}\frac{\delta(p_{//})}{\mu^2+p_\bot^2+p_{//}^2}=
$$
$$
=-\frac{g^2}{4(2\pi)^3 s}\int d^2 p_\bot \frac{e^{i(p_\bot
r_\bot)}}{\mu^2+p_\bot^2} =-\frac{g^2}{4(2\pi)^2 s}K_0\left(\mu
|r_\bot|\right),\eqno(B.2)
$$

with $K_0\left(\mu |r_\bot|\right)=\frac{1}{2\pi}\int d^2 p_\bot
\frac{e^{i(p_\bot r_\bot)}}{\mu^2+p_\bot^2}$ is the MacDonald
function of zeroth order. \\

 Considering the integral

$$
I_2=\int d^2r_\bot e^{i\Delta_\bot
r_\bot}K_0(\mu|r_\bot|).\eqno(B.3)
$$

we have

$$
I_2=(2\pi)\int d |r_\bot||r_\bot| J_{(0)}(\Delta_\bot
|r_\bot|)K_0\left(\mu |r_\bot|\right)
$$
$$
=(2\pi)\frac{1}{\Delta_\bot^2 +\mu^2}=\frac{2\pi}{\mu^2-t}.
\eqno(B.4)
$$

Considering the integral

$$
I_3=\int d^2r_\bot e^{i\Delta_\bot
r_\bot}K_0^2(\mu|r_\bot|).\eqno(B.5)
$$
$$
=\int d^2 r_\bot e^{i\Delta_\bot r_\bot}\left(\frac{1}{2\pi}\int
d^2q \frac{e^{i\textbf{qr}_\bot}}{q^2+\mu^2}\right)K_0\left(\mu
|r_\bot|\right)=
$$
$$
=\frac{1}{2\pi}\int d^2q \frac{1}{q^2+\mu^2}\int d^2 r_\bot
e^{i\left(q+\Delta_\bot\right) r_\bot}K_0\left(\mu|r_\bot|\right)
$$
$$=\frac{1}{2\pi}(2\pi)\int d^2q
\frac{1}{q^2+\mu^2}\frac{1}{\left(q+\Delta_\bot\right)^2 +\mu^2},
$$

here, we used the result of integral that obtained from calculating
$I_2$. \\

Using method of Feynman parameter integral
$\frac{1}{ab}=\int_0^1\frac{dx}{[ax+b(1-x)]^2}$, we have

$$
I_3=\int_0^1 dx\int d^2q \frac{1}{\left\{(q^2+\mu^2)x
+\left[\left(q+\Delta_\bot\right)^2+\mu^2\right](1-x)\right\}^2}=
$$
$$
=\int_0^1 dx \int d^2q\frac{1}{\left[q^2+2q\Delta_\bot(1-x)
+\Delta_\bot^2(1-x)+\mu^2\right]^2}=
$$
$$
=\int_0^1 dx
\frac{i(-\pi)\Gamma(1)}{\left[\Delta_\bot^2(1-x)+\mu^2-\Delta_\bot^2(1-x)^2\right]\Gamma(2)}
$$
$$
=(-i\pi)\int_0^1 dx\frac{1}{\left[\mu^2+\Delta_\bot^2x(1-x)\right]}
 =(-i\pi)\int_0^1\frac{dx}{\left[\mu^2-tx(1-x)\right]}=
$$
$$
=(-i\pi)\times\frac{1}{t\sqrt{1-\frac{4\mu^2}{t}}}
ln\frac{1-\sqrt{1-\frac{4\mu^2}{t}}}{1+\sqrt{1-\frac{4\mu^2}{t}}}
\equiv(-i\pi)\times F_{1}(t). \eqno(B.6)
$$

Finally, we calculate the integral

$$
I_4=\int d^2r_\bot e^{i\Delta_\bot r_\bot}K_0^3(\mu|r_\bot|)=
$$
$$
=\int d^2r_\bot e^{i\Delta_\bot r_\bot}\left(\frac{1}{2\pi}\int
d^2q_1\frac{e^{iq_1r_\bot}}{q_1^2+\mu^2}\right)\left(\frac{1}{2\pi}\int
d^2q_2\frac{e^{iq_2r_\bot}}{q_2^2+\mu^2}\right)K_0(\mu
\left|r_\bot\right|)=
$$
$$
=\frac{1}{(2\pi)^2}\int\frac{d^2q_1d^2q_2}{(q_1^2+\mu^2)(q_2^2+\mu^2)}
\int d^2x_\bot\exp\left[i(q_1+q_2+\Delta_\bot)x_\bot\right]K_0(\mu
\left|r_\bot\right|)=
$$
$$
=\frac{1}{(2\pi)^2}\int d^2 q_1 d^2 q_2
\frac{1}{(q_1^2+\mu^2)(q_2^2+\mu^2)\left[(q_1+q_2+\Delta_\bot)^2+\mu^2\right]}.\eqno(B.7)
$$

Applying the result that we obtained when calculating $I_3$, we have

$$
\int d^2q_1
\frac{1}{(q_1^2+\mu^2)\left[(q_1+q_2+\Delta_\bot)^2+\mu^2\right]}
=(-i\pi)\int_0^1 dx\frac{1}{\left[\mu^2+(q_2+\Delta_\bot)^2
x(1-x)\right]},
$$
so
$$
 I_4=\frac{1}{(2\pi)^2}(-i\pi)\int_0^1 dx\int d^2q_2
 \frac{1}{(q_2^2+\mu^2)\left[\mu^2+(q_2+\Delta_\bot)^2
x(1-x)\right]}\eqno(B.8)$$

From method of Feynman parameter integral, again, we obtain

$$I_4=-\frac{i}{(4\pi)}\int_0^1\frac{dx}{x(1-x)}\int_0^1dy\int
d^2q_2\frac{1}{\left\{\left[(q_2+\Delta_\bot)^2+B\right]y+(q_2^2+\mu^2)(1-y)\right\}^2}=$$
$$=-\frac{i}{(4\pi)}\int_0^1\frac{dx}{x(1-x)}\int_0^1dy\int
d^2q_2\frac{1}{\left(q_2^2+2q_2\Delta_\bot y+C\right)^2}=$$
$$=-\frac{i}{(4\pi)}(-i\pi)\int_0^1\frac{dx}{x(1-x)}\int_0^1dy
\frac{1}{\left[C-(\Delta_\bot y)^2\right]}=$$
$$=-\frac{1}{4}\int_0^1\frac{dx}{x(1-x)}\int_0^1dy
\frac{1}{\left[C-(\Delta_\bot y)^2\right]}.)$$

where

$$B=\frac{\mu^2}{x(1-x)};C=(\Delta_\bot^2+B)y+\mu^2(1-y)
=\left[\frac{\mu^2}{x(1-x)}-t\right]y+\mu^2(1-y),\eqno(B.9)$$
then
$$I_4=-\frac{1}{4}\int_0^1\frac{dx}{x(1-x)}\int_0^1dy
\frac{1}{\left[\frac{\mu^2}{x(1-x)}-t\right]y+\mu^2(1-y)+ty^2}=$$
$$=-\frac{1}{4}\int_0^1 dy\int_0^1 dx \frac{1}{-(1-y)(ty-\mu^2)x(1-x)+\mu^2}$$
$$=-\frac{1}{4}\int_0^1 dy\int_0^1 dx \frac{1}{Dx^2-Dx+\mu^2}=$$
$$=-\frac{1}{4}\int_0^1
\frac{dy}{D}\int_0^1\frac{dx}{x^2-x+\frac{\mu^2}{D}}=-\frac{1}{4}\int_0^1
\frac{dy}{D}\int_0^1\frac{dx}{(x-x_1)(x-x_2)}=$$
$$=-\frac{1}{4}\int_0^1\frac{dy}{D}\int_0^1
dx\left[\frac{1}{x-x_1}-\frac{1}{x-x_2}\right]\frac{1}{x_1-x_2}=$$
$$
=-\frac{1}{4}\int_0^1\frac{dy}{D}\frac{1}{x_1-x_2}ln\left|\frac{(1-x_1)x_2}
{(1-x_2)x_1}\right|,\eqno(B.10)$$

here, $D= -(1-y)(ty-\mu^2)=ty^2+(\mu^2-t)y+\mu^2$ and $x_1;x_2$ are
roots of equation $
x^2 -x+\frac{\mu^2}{D}=0 $.\\

Noting that: $x_1+x_2=1\Rightarrow 1-x_1=x_2;1-x_2=x_1$, and:
$$x_1-x_2=\sqrt{1-\frac{4\mu^2}{D}}\approx1-\frac{2\mu^2}{D},\eqno(B.11)$$
hence

$$ln\left|\frac{(1-x_1)x_2}{(1-x_2)x_1}\right|=ln\left|\frac{x_2^2}{x_1^2}\right|
=2ln\left|\frac{1-\sqrt{1-\frac{4\mu^2}{D}}}{1+\sqrt{1-\frac{4\mu^2}{D}}}\right|\approx$$
$$\approx2ln\left|\frac{1-\left(1-\frac{2\mu^2}{D}\right)}{1+\left(1-\frac{2\mu^2}{D}\right)}\right|
=2ln\left|\frac{\mu^2}{D-\mu^2}\right|.\eqno(B.12)$$

so that
$$I_4=-\frac{1}{2}\int_0^1 dy\frac{1}{D-2\mu^2}ln\left|\frac{\mu^2}{D-\mu^2}\right|=$$
$$=-\frac{1}{2}\int_0^1 dy\frac{1}{(ty+\mu^2)(y-1)}ln\left|\frac{\mu^2}{y(ty+\mu^2-t)}\right|\equiv-\frac{1}{2}F_2(t)\eqno(B.13)$$

\end{document}